\documentclass{emulateapj}
\usepackage{apjfonts}
\usepackage{lscape}

\begin{document}


\title{Demographics of Bulge Types within 11~Mpc and Implications for
  Galaxy Evolution}

\shorttitle{Demographics of Bulges}
\shortauthors{Fisher \& Drory}
      

\author{David~B.~Fisher\altaffilmark{1} } 
\affil{Laboratory of Millimeter Astronomy, University of
  Maryland, College Park, MD 29742}
\altaffiltext{1}{Department of Astronomy, The University of Texas at Austin,\\
  1 University Station C1400, Austin, Texas 78712}
\email{dbfisher@astro.umd.edu}

\author{Niv~Drory\altaffilmark{2}}
\affil{Instituto de Astronom\'ia, Universidad Nacional Aut\'onoma de M\'exico, A.P.\ 70-264, 04510 M\'exico, D.F., M\'exico}
\altaffiltext{2}{Max-Planck-Institut f\"ur
  Extraterrestrische Physik, Giessenbachstra\ss e, 85748 Garching, Germany}

\slugcomment{Submitted to ApJL}


\begin{abstract}
  We present an inventory of galaxy bulge types (elliptical galaxy,
  classical bulge, pseudobulge, and bulgeless galaxy) in a
  volume-limited sample within the local 11~Mpc volume using Spitzer
  3.6~$\mu$m and HST data. We find that whether counting by number,
  star formation rate, or stellar mass, the dominant galaxy type in
  the local universe has pure disk characteristics (either hosting a
  pseudobulge or being bulgeless). Galaxies that contain either a
  pseudobulge or no bulge combine to account for over 80\% of the
  number of galaxies above a stellar mass of
  $10^{9}$~M$_{\odot}$. Classical bulges and elliptical galaxies
  account for $\sim$1/4, and disks for $\sim$3/4 of the stellar mass
  in the local 11~Mpc. About 2/3 of all star formation in the local
  volume takes place in galaxies with pseudobulges.  Looking at the
  fraction of galaxies with different bulge types as a function of
  stellar mass, we find that the frequency of classical bulges
  strongly increases with stellar mass, and comes to dominate above
  $10^{10.5}$~M$_{\odot}$. Galaxies with pseudobulges dominate at
  $10^{9.5}$-$10^{10.5}$~M$_{\odot}$. Yet lower-mass galaxies are most
  likely to be bulgeless. If pseudobulges are not a product of
  mergers, then the frequency of pseudobulges in the local universe
  poses a challenge for galaxy evolution models.
\end{abstract}



\section{Introduction}\label{sec:intro}

Hierarchical galaxy evolution models \citep[e.g.][]{whiterees1978,
  Cole+1994} rely on the assumption that bulge-to-total ratios
increase directly, and exclusively, from merging (reviewed in
\citealp{Baugh2006}). This has been justified by the ability of
simulations of mergers to reproduce properties of ellipticals
\citep[e.g.][]{cox2006,naab2006}, and the extrapolation motivated by
observations of notable galaxies (e.g.\ M~31) that bulges are similar
to ellipticals.

Yet, there is a dichotomy in the properties of bulges and possibly in
their formation mechanisms. Some bulges are similar to elliptical
galaxies ({\em classical bulges}), other bulges resemble disks ({\em
  pseudobulges}; for reviews see \citealp{kk04,combes2009}). Cursory
analysis suggests that simulations producing bulge-disk galaxies
\citep[e.g.][]{governato2009} are likely not making pseudobulges.

Many authors propose that disk-like bulges form through internal
secular evolution of the disk \citep[for reviews see][]{kk04,athan05}.
\cite{fisherdrory2008} and \cite{fisherdrory2010} show that
pseudobulges have S\'ersic index $n<2$ and do not follow projections
of the fundamental plane of elliptical galaxies, adding evidence that
pseudobulges are physically different from classical bulges (which
have $n>2$) and ellipticals.  \cite{fdf2009} find that pseudobulges
typically have high enough star formation rates (SFR) to have built
their stellar mass within the typical lifetime of a disk. Furthermore,
correlations between bulge and disk properties such as stellar age
\citep{peletier1996} and radial size \citep{fisherdrory2008} may
result from a formative link between pseudobulges and their
surrounding disk. Indeed, \cite{fisherdrory2010} find that the only
property that correlates with the half-light radius of pseudobulges is
the outer disk scale length.

\cite{heller2007} show that significant gaseous inflow occurs across
the central kpc during bar lifetimes. \cite{bureau1999} show evidence
that boxy/peanut shaped bulges are the result of bar-buckling in
disks.  Boxy bulges are found in over 40\% of edge on galaxies
\cite{lutticke2000}, thus implying that a significant number of bulges
may owe their origin to disk phenomena.  We caution that secular
evolution and accretion/merging are not mutually exclusive
\citep{bournaud2002}. \cite{fisherdrory2010} find that some
pseudobulges ($\lesssim13$\% of their sample) could house a small
classical bulge, and still maintain a low S\'ersic index.

How common are pseudobulges? \cite{droryfisher2007} find that
classical bulges are exclusively found in red-sequence galaxies, and
imply that pseudobulges are at least as common as blue, Sa-Sc galaxies.
\cite{kormendy2010} find that in the local 8~Mpc, 11 of 19 galaxies
with $V_{c}>150$~km~s$^{-1}$ show no evidence for a classical bulge;
however, this is a small sample that does not allow to study the mass
dependence of the frequency of pseudobulges. \cite{weinzirl2009} show
that traditional semi-analytic models of galaxy formation cannot
account for the observed number of small bulges. This discrepancy may
be a manifestation of the bulge dichotomy, since pseudobulges are more
likely to be in low $B/T$ galaxies \citep{fdf2009}. However, many
pseudobulges have $B/T>0.2$ \citep{fisherdrory2008,fisherdrory2010}.

In this letter, we study the abundance of pseudobulges and classical
bulges in the local universe. We determine bulge-types on a sample
including all non-edge-on galaxies having $B<15$ within 11~Mpc ($M_B <
-15.2$) and estimate the dependence of pseudobulge frequency on galaxy
mass and SFR.

\section{Methods}

We select a representative volume-limited sample of non-edge-on
($i<80\degr$) galaxies within 11~Mpc from the \cite{11hugs} survey,
complete for spirals to $B=15$~mag (corresponding to $M_B=-15.2$). We
require Galactic latitude $|b|>20\degr$. We take $B_{T}$ values from
\cite{rc3} and HyperLEDA\footnote{\tt http://leda.univ-lyon1.fr/} in
order of preference. Since the \cite{11hugs} sample does not cover
early-type galaxies, we add these from \cite{tonry2001},
\cite{tully1988}, and HyperLEDA using the same magnitude and Galactic
latitude cuts. Because bulge diagnosis is not reliable on edge-on
galaxies, we exclude disks with inclination greater than 80\degr. This
selection may overemphasize the number of E-galaxies by 10\% as they
are not flattened. We adopt distances from \cite{11hugs} augmenting
missing data from \cite{tonry2001}, \cite{tully2009}, and
\cite{tully1988}.  Magnitudes and colors are corrected for extinction
\citep{schlegel} and galaxy inclination in the usual manner. The final
sample contains 320 galaxies. The full sample and measured quantities
are listed in Table~1.

We decompose the major-axis near-IR surface brightness (SB) profile of
97 bright ($M_B<-16$~mag) and non Sm/Irr galaxies at 3.6~$\mu$m (2MASS
$K$-band for 6 galaxies) into a S\'ersic-function bulge and
exponential outer disk. Non-exponential disk components (e.g.\ bars
and rings) are masked. Most of our decompositions are taken from
\cite{fisherdrory2010}. This analysis has been used in many
publications including \cite{fisherdrory2008,kfcb,fisherdrory2010}.
The S\'ersic index, $n$ is used to diagnose bulges into pseudo-
($n<2$) and classical ($n>2$) bulges (see \citealp{fisherdrory2008}
for a discussion).  For those bulges with $n\sim2$ we supplement bulge
identification with nuclear morphology from HST images.  Ellipticals
are assigned $B/T = 1$. Galaxies in which the decomposition yields
$B/T < 0.01$ are assigned $B/T=0$ and are called ``bulgeless''.  We
determine total luminosity by integrating the near-IR SB profile and
convert to stellar mass using RC3 $B-V$ color as described in
\cite{fdf2009}, following \cite{belldejong2001}. Seven bright galaxies
have no $B-V$ recorded and for these we substitute the average color
of their Hubble type.

We assume that the 223 faint ($M_B>-16$~mag) or Sm/Irr galaxies in our
sample are bulgeless. Most have no usable near-IR data; we therefore
use $M_B$ in conjunction with $B-V$ to determine stellar mass.  123
do not have a measurement of $B-V$ and we again use the mean color of
their Hubble type instead. For a handful of galaxies we test this
against masses determined from near-IR flux, finding good agreement.


Available means of measuring SFR in our sample include GALEX FUV
luminosity, H$\alpha$ luminosity, and 24~$\mu$m dust emission; linear
combination of either H$\alpha$ or UV (unobscured light) with
24~$\mu$m (extincted light) being most robust.

In galaxies fainter than $M_B=-16$~mag, \cite{lee2009} find that the
UV SFR is systematically higher than that from other tracers, possibly
due to differences in the stellar IMF.  Therefore, we calculate the
SFR from H$\alpha$ and FUV according to \cite{kennicutt98araa} and
take the higher of the two values.

For 78 of the 97 bright galaxies, we measure the total SFR and the SFR
within the central 1~kpc by linearly combining the 24~$\mu$m and GALEX
FUV data \citep{leroy2008,fdf2009}, $SFR=a \times [L(\mathrm{FUV}) + b
\times L(24)]$, where $a$ and $b$ are constants calibrated against
\cite{kennicutt2009}. The 19 remaining galaxies lack GALEX data. For 6
of these, we measure SFR by linearly combining 24~$\mu$m with total
H$\alpha$ luminosity of \cite{11hugs}, $SFR\propto
L(\mathrm{H}\alpha)+a_{24}\times L(24)$, according to
\cite{kennicutt2009}. The SFR of the central 1~kpc is measured with
24~$\mu$m alone as $SFR\propto L(24)^{0.885}$ following
\cite{calzettietal2007}. Four galaxies have data at 24~$\mu$m only;
for these we follow \cite{fdf2009}. One galaxy has only H$\alpha$ and
one has only UV data; there we use the single band flux ($SFR\propto
L(\mathrm{H}\alpha)$ or $SFR\propto L(\mathrm{FUV})$) following
\cite{kennicutt98araa} and we cannot measure the luminosity of the
central kpc. Finally, 7 of the bright galaxies have no data available
for measuring SFR. The method applied to calculate SFR for each galaxy is
noted in Table~1.

Uncertainties in stellar mass and SFR are dominated by the
scatter in the calibration of measured fluxes to physical
quantities. The calibration error for stellar mass is 0.12~dex for
near-IR flux and 0.16~dex for $M_B$. The calibration error for SFR is
roughly 15\% for data combining H$\alpha$+24~$\mu$m and 
FUV+24~$\mu$m, and is closer to 20\% for data using FUV or H$\alpha$.

\section{Results} 

Before discussing our results, we call attention to the environmental
bias inherent in studying galaxies in the local 11-Mpc volume due to
the low density of that region \citep[reviewed in][]{peebles2010}. For
comparison, \cite{kfcb} finds that 2/3 of all stellar mass in the
Virgo cluster is in elliptical galaxies alone.

{\bf Bulge number statistics:} Galaxies with either a pseudobulge or
no bulge are the most common among bright galaxies. Restricting
ourselves to galaxies more massive than $10^9$~M$_{\odot}$, we find
that only 17$\pm$10\% are galaxies with an observed classical bulge
(including elliptical galaxies), 45$\pm$12\% are galaxies with
pseudobulges, 35$\pm$12\% are are disk galaxies with $B/T<0.01$, and
under 3\% are galaxies currently undergoing major merging (NGC~4490,
NGC~1487, NGC~2537). Quoted errors are Poisson uncertainties. Dwarf
and Irregular galaxies comprise $\sim$70\% of all galaxies having
stellar mass lower than $10^9$~M$_{\odot}$ within 11~Mpc. However,
they only account for $\sim$2\% of the stellar mass in the same
volume.

\begin{figure}[t]
\begin{center}
\includegraphics[width=0.49\textwidth]{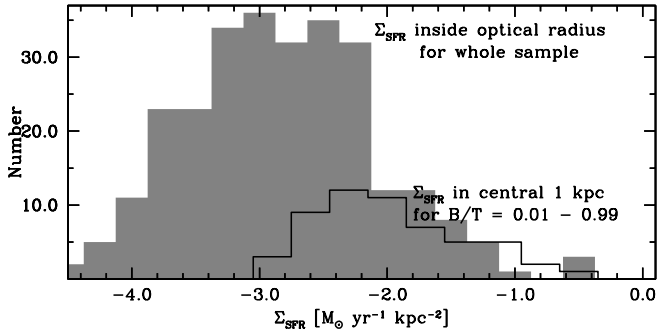}
\end{center}
\caption{The distribution of SFR density $\Sigma_{\mathrm{SF}}(r<1~$kpc$)$ 
for bulges (black line). For comparison, we also show the SFR density of 
entire galaxies ($\Sigma_{\mathrm{SF}}$(total); grey shaded region).\label{fig:sfr}}
\end{figure}

{\bf Star formation in bulge-disk galaxies:} 61\% of the star
formation (SF) in the local 11~Mpc is in galaxies with pseudobulges. A
non-negligible 13\% of the total SF in our volume occurs in the
central kpc of bulge-disk galaxies. Fig.~\ref{fig:sfr} shows the
distribution of SFR surface densities ($\Sigma_{\mathrm{SF}}$) of
entire galaxies and the central kpc of bulge-disk galaxies. It is
clear that high $\Sigma_{\mathrm{SF}}$ in the central kpc of
bulge-disk galaxies is extremely common when compared to global SFR
densities. In our sample, we find that 46$\pm$9\% of galaxies with
bulge-to-total ratios in the range $0.01 \leq B/T < 1$ have
$\Sigma_{\mathrm{SF}}>10^{-2}$M$_{\odot}$~yr$^{-1}$kpc$^{-2}$; only
33$\pm$9\% of entire galaxies have
$\Sigma_{\mathrm{SF}}>10^{-2}$M$_{\odot}$~yr$^{-1}$kpc$^{-2}$ inside
the optical radius.  In the bulge sample, 11 bulges do not have data
to determine the SFR. If these have low SFR, the fraction of bulges
with high $\Sigma_{\mathrm{SF}}$ decreases to 35$\pm$10\%.

\begin{figure}[t]
\begin{center}
\includegraphics[width=0.45\textwidth]{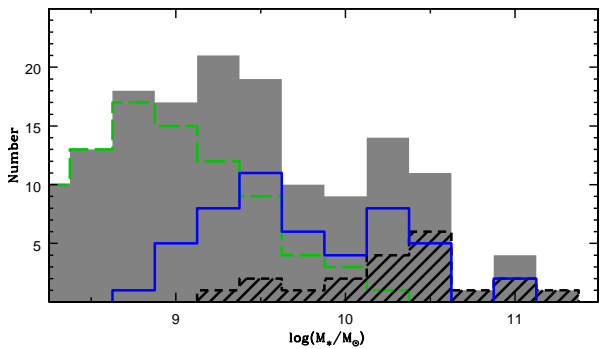}
\end{center}
\caption{The distribution of galaxy stellar mass in galaxies with
  pseudobulges (blue line), elliptical galaxies and galaxies with
  classical bulges (short-dashed line), bulgeless galaxies (green
  long-dashed line) and the full sample (grey shaded region). Note the
  full range of stellar masses in our sample is not
  shown. \label{fig:mass}}
\end{figure}

{\bf Stellar masses:} Fig.~\ref{fig:mass} shows the stellar-mass
distribution of galaxies with pseudobulges, classical bulges and
ellipticals (combined), bulgeless galaxies, and the whole sample.
Bulgeless galaxies tend to be lower in mass, and dominate the
distribution up to $M_{*}\sim 10^{9.5}$~M$_{\odot}$. Pseudobulges
dominate intermediate mass range from, $M_{*}\gtrsim \times10^{9.5}$
to $10^{10.5}$~M$_{\odot}$. Classical bulges tend to be in more
massive galaxies. Galaxies with either a pseudobulge or no bulge
combine to account for 56$\pm$12\% of the stellar mass of galaxies
within 11~Mpc.  Finally, we calculate the total mass in classical
bulges by using the $B/T$ from the bulge-disk decompositions. These
values should be treated as estimates, since they assume the same
$M/L$ for both the bulge and disk, hence likely underestimating the
classical bulge mass. Classical bulges and E galaxies account for
$\sim$1/4 of the stellar mass in the local 11~Mpc, disks account for
$\sim 3/4$ of the stellar mass.

\section{Discussion}

We show that galaxies with pseudobulges are the most common type of
bright galaxy in the local 11~Mpc volume. The set of galaxies
including pseudobulge and bulgeless galaxies account for just over 1/2
of the mass in stars in the local volume. Roughly $2/3$ of new stars
are made in galaxies with pseudobulges. Whether counting by number,
mass, or by present-day star formation, the dominant mode of galaxy
evolution in the present day local universe is that which occurs in
galaxies without classical bulges. These results are therefore in
agreement with the observed correlation of bulge type with galaxy
properties such as color \citep{droryfisher2007}.

We find that classical bulges and elliptical galaxies combined account
for $\sim$1/4 of the stellar mass within 11~Mpc. Therefore, 3/4 of the
stellar mass in the local 11~Mpc is in disks (combining all mass in
pseudobulges, disks around classical bulges and pseudobulges, and
bulgeless galaxies).  Recall that in cluster environments, 2/3 of the
stellar mass is in elliptical galaxies alone \citep{kfcb}. Thus the
process driving the distribution of bulge types appears to be a strong
function of environment.

We show that in the majority of bulge-disk galaxies, the central kpc
has high SFR surface density (compared to the SFR density for entire
galaxies).  If a merger drives enhanced SFR for 1~Gyr \citep{cox2008},
and if fewer than 10\% of giant galaxies experience merging each Gyr
\citep[e.g.][]{jogeeetal2009J}, then episodic SF can not account for
the frequency of enhanced SF observed in our sample, and thus the SF
in the centers of (pseudo)bulges is not likely episodic or merger
driven. The frequency of enhanced SF in bulges is thus further
evidence that bulges are generating new stars through long term,
non-episodic processes.

\begin{figure}
\begin{center}
\includegraphics[width=0.49\textwidth]{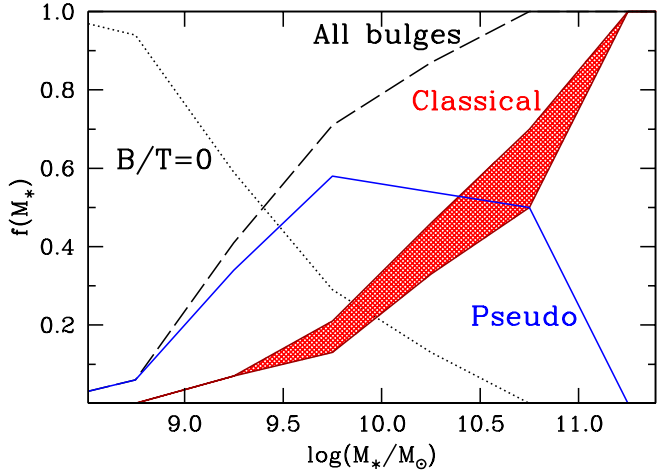}
\end{center}
\caption{The relative number of galaxies with
  classical bulges and elliptical galaxies (red lines), galaxies containing pseudobulges (blue line), 
  all disk-bulge galaxies (black dashed line), and bulgeless galaxies (black dotted line) as a function of
  galaxy stellar mass.
  \label{fig:freq}}
\end{figure}

Finally, in Fig.~\ref{fig:freq}, we estimate the relative frequency of
classical bulges (including elliptical galaxies), pseudobulges, all
bulges, and galaxies with no bulge within 11~Mpc as a function of
galaxy stellar mass. To account for the possibility of composite
systems, we estimate an upper bound for the frequency of classical
bulges: we include all those bulges that satisfy the criteria to be
called classical and elliptical galaxies, add all galaxies presently
in strong interactions (NGC~4490, NGC~1487, NGC~2537 \& NGC~5194A \&
B), and we estimate the possible number of galaxies with composite
(pseudo+classical) bulges. \cite{fisherdrory2010} find that models of
bulges in which the total bulge light is composed of a high and low
S\'ersic index component are not inconsistent with decompositions of
real bright low-specific-SFR pseudobulges. Consistent with these
results, we select all pseudobulges with stellar mass
$M_{\mathrm{pseudo}}>10^{9}$~M$_{\odot}$ and specific SFR
$<0.03$~Gyr$^{-1}$ as candidate composite bulges.  For the interacting
galaxies we make the assumption that a merger will result in an
elliptical and thus $B/T=1$.

Fig.~\ref{fig:freq} shows that the frequency of pseudobulges and
classical bulges in the local universe is strongly dependent on galaxy
mass.  Pure disk galaxies and those galaxies with pseudobulges are the
most common type of galaxy for stellar mass $M_{*} \leq
10^{10}$~M$_{\odot}$. Elliptical galaxies and galaxies with classical
bulges are the majority of galaxies with $M_{*} \geq
10^{10.5}$~M$_{\odot}$. However, since galaxies with $M_{*} \geq
10^{10.5}$~M$_{\odot}$ only make up 4\% of bright galaxies in the
local volume, galaxies with pseudobulges and those with no bulge
remain the dominant type of bright galaxy by number.  Dynamical
evidence suggest that the Milky Way (not included in the sample) does
not contain a classical bulge \citep{shen2010}, its stellar mass
places it right at the transition, $M_{*,MW}
\sim10^{10.5}$~M$_{\odot}$. Therefore, the massive galaxies in the
Local Group comprise a pseudobulge galaxy (Milky Way), a classical
bulge galaxy (M~31), and a bulgeless disk galaxy (M~33).

The simulation of the evolution of galaxies in a $\Lambda$CDM-universe
by \cite{croft2009} provides a good model for comparison. As is
normally the case, in this simulation $B/T$ is only increased through
the merging process. They find that in massive galaxies ($\gtrsim
10^9$~M$_{\odot}$) located in low density environments (i.e.\ field
galaxies), 40-50\% are bulge-dominated ($B/T$>80\%).  In the local
11~Mpc only 23$\pm$5\% of galaxies contain classical bulges at any
bulge-to-total ratio (including ellipticals and ongoing mergers), and
only 5\% have $B/T$>80\%.  If we assume that the simulation in
\cite{croft2009} only produces classical bulges, then the number of
classical bulges in the local universe is much smaller than in a
typical galaxy evolution simulation.

Recently, \cite{hopkins2009_gas} show that if $B/T$ is a function of
both merger mass-ratio and gas fraction in the progenitive merger,
then the distribution of $B/T$ for all galaxies is recovered. However,
this agreement relies entirely on the interpretation of pseudobulges
as merger products (contrary to observational evidence). In our
sample, the fraction of classical-bulge light in galaxies less massive
than $M_{*}\sim 10^{10}$~M$_{\odot}$ is very low, $B/T\lesssim 5\%$.
If pseudobulges are not merger products, but rather disk components,
then models continue to produce too much mass in bulges.

We conclude that pseudobulges and internal bulge growth through SF is
present in the majority of giant disk galaxies in the local 11~Mpc
volume. If we make the assumption that pseudobulges are not direct
merger products, then the number of pseudobulges poses a challenge for
models of galaxy evolution.  Given that very old stellar populations
are commonly observed in spiral galaxies \citep{macarthur2009},
holding off disk galaxy formation until lower redshifts does not
appear to be the solution. The problem is that, {\em as we understand
  them now}, mergers in recent epochs are likely to increase $B/T$ and
heat the disk thereby reducing the secular inward flow of gas in
disks, and possibly destroying a pre-existing pseudobulge in a disk
galaxy.  Therefore, either the merging process does not disrupt disks
as easily as simple calculations suggest
\citep[see][]{hopkins2009_disk_mergers,moster2010}, or there are fewer
galaxy mergers in recent epochs in the universe than simulations
suggest.
 

\acknowledgments

DBF acknowledges support from University of Maryland, as well as
J~Kormendy and the University of Texas at Austin. ND and DBF thank the
Max-Planck Society for support during this project. We also wish to
thank Shardha Jogee, Karl Gebhardt, Neal Evans, John Kormendy and Ralf
Bender for their helpful comments and support during the writing of
this work.  This work is based on observations made with the Spitzer
Space Telescope, which is operated by the Jet Propulsion Laboratory,
California Institute of Technology under a contract with NASA. Support
for this work was provided by NASA through an award issued by
JPL/Caltech. DBF acknowledges support by the National Science
Foundation under grant AST 06-07490.  Some of the data presented in
this paper were obtained from the Multi-mission Archive at the Space
Telescope Science Institute (MAST). STScI is operated by the
Association of Universities for Research in Astronomy, Inc., under
NASA contract NAS5-26555. Support for MAST for non-HST data is
provided by the NASA Office of Space Science via grant NAG5-7584 and
by other grants and contracts.

\bibliography{thesis} \bibliographystyle{apj}


\clearpage 
\begin{center}
\LongTables
\begin{landscape} 
\begin{deluxetable}{lcccccccccc}
  \tablewidth{0pt} \tablecaption{Sample Data}
\tablehead{\colhead{Galaxy} & \colhead{Category}  & \colhead{T} &
  \colhead{Dist.} & \colhead{M$_B$} &
  \colhead{log(M$_{*,{\mathrm{total}}}$) } &
    \colhead{log($\psi_{\mathrm{total}}$)}  & \colhead{SFR} &
    \colhead{B/T} & \colhead{S\'ersic} &
    \colhead{log($\psi_{\mathrm{1~kpc}}$)} \\
\colhead{Name } & \colhead{\tablenotemark{(a)}} & \colhead{} &\colhead{(Mpc)} &
\colhead{(mag)} &\colhead{(M$_{\odot}$)} & \colhead{(M$_{\odot}$
  yr$^{-1}$) } & \colhead{Method} & \colhead{ } & \colhead{index} &\colhead{(M$_{\odot}$
  yr$^{-1}$) }  }
\startdata
NGC6744 & C & 4 & 9.4 & -21.2 & 10.36 & -0.27 & UV,24 & 0.15 & 3.2 $\pm$ 1.1 & -2.33 \\
NGC0224 & C & 3 & 0.8 & -21.2 & 10.62 & -1.85 & 24 & 0.48 & 2.1 $\pm$ 0.5 & -2.04 \\
NGC5194 & P\tablenotemark{(b)} & 4 & 8.0 & -21.2 & 10.93 & 0.35 & UV,24 & 0.11 & 0.5 $\pm$ 0.3 & -0.83 \\
NGC4594 & C & 1 & 9.3 & -21.1 & 10.96 & -0.70 & UV,24 & 0.51 & 6.2 $\pm$ 0.6 & -1.51 \\
NGC4258 & C\tablenotemark{(c)} & 4 & 8.0 & -21.0 & 10.49 & ... & ... & 0.11 & 2.8 $\pm$ 0.6 & ... \\
NGC4490 & M & 7 & 8.0 & -20.9 & 10.14 & -0.36 & UV,24 & 1.00 & ...   & ... \\
NGC3627 & P & 3 & 10.1 & -20.9 & 10.54 & 0.05 & UV,24 & 0.10 & 1.4 $\pm$ 0.7 & -0.91 \\
NGC2903 & P & 4 & 8.9 & -20.9 & 10.29 & 0.04 & UV,24 & 0.10 & 0.5 $\pm$ 0.1 & -0.39 \\
NGC0253 & P & 5 & 3.2 & -20.9 & 10.62 & 0.22 & UV,24 & 0.05 & 1.5 $\pm$ 0.6 & -0.20 \\
NGC5457 & P & 6 & 6.7 & -20.8 & 10.24 & 0.33 & UV,24 & 0.02 & 1.5 $\pm$ 1.8 & -1.51 \\
NGC5236 & P & 5 & 4.5 & -20.7 & 10.22 & 0.20 & UV,24 & 0.09 & 0.4 $\pm$ 0.1 & -0.15 \\
NGC3031 & C & 2 & 3.6 & -20.7 & 10.66 & -0.85 & UV,24 & 0.37 & 3.9 $\pm$ 0.5 & -1.67 \\
NGC4826 & C\tablenotemark{(c)} & 2 & 7.5 & -20.6 & 10.56 & -0.48 & UV,24 & 0.29 & 3.6 $\pm$ 0.7 & -0.72 \\
NGC1291 & C & 0 & 9.4 & -20.5 & 10.98 & -0.34 & UV,24 & 0.47 & 2.7 $\pm$ 0.8 & -0.92 \\
NGC5055 & P & 4 & 7.5 & -20.5 & 10.48 & -0.10 & UV,24 & 0.19 & 1.3 $\pm$ 1.4 & -1.19 \\
NGC3368 & P & 2 & 10.5 & -20.4 & 10.49 & -0.69 & UV,24 & 0.26 & 1.6 $\pm$ 0.4 & -1.39 \\
NGC4559 & nb/d & 6 & 9.7 & -20.3 & 10.34 & -0.15 & UV,24 & 0.00 & ...   & ... \\
NGC3521 & C & 4 & 8.0 & -20.3 & 10.51 & -0.13 & UV,24 & 0.12 & 2.6 $\pm$ 1.6 & -1.20 \\
NGC3556 & P\tablenotemark{(d)}  & 6 & 10.4 & -20.1 & 10.23 & ... & ... & 0.21 & 2.1 $\pm$ 1.1 & ... \\
NGC3034 & nb/d & 7 & 3.5 & -20.1 & 10.04 & -0.03 & Ha & 0.00 & ...   & ... \\
NGC3621 & P\tablenotemark{(d)} & 7 & 6.6 & -20.0 & 10.37 & -0.40 & UV,24 & 0.01 & 2.8 $\pm$ 1.0 & -1.54 \\
NGC0925 & P & 7 & 9.2 & -20.0 & 9.53 & -0.45 & UV,24 & 0.07 & 0.7 $\pm$ 0.6 & -1.47 \\
NGC3379 & E & -5 & 10.5 & -20.0 & 11.18 & -1.53 & UV,24 & 1.00 & ...   & ... \\
NGC0628 & P & 5 & 7.3 & -20.0 & 9.87 & -0.35 & UV,24 & 0.08 & 1.6 $\pm$ 0.3 & -1.76 \\
NGC3351 & P & 3 & 10.0 & -19.9 & 10.29 & -0.16 & UV,24 & 0.16 & 1.5 $\pm$ 0.4 & -0.45 \\
NGC4736 & P & 2 & 4.7 & -19.8 & 10.27 & -0.59 & UV,24 & 0.36 & 1.3 $\pm$ 0.4 & -0.99 \\
NGC3623 & P & 1 & 7.3 & -19.7 & 10.95 & -0.86 & UV,24 & 0.16 & 1.8 $\pm$ 0.8 & -2.15 \\
NGC3675 & P & 3 & 10.6 & -19.6 & 10.32 & -0.53 & UV,24 & 0.14 & 1.6 $\pm$ 1.3 & -1.23 \\
NGC4096 & P & 5 & 8.3 & -19.6 & 10.12 & -0.75 & UV,24 & 0.08 & 0.8 $\pm$ 0.4 & -1.57 \\
NGC2403 & P & 6 & 3.2 & -19.4 & 9.43 & -0.65 & UV,24 & 0.07 & 0.7 $\pm$ 0.7 & -1.85 \\
NGC5195 & P\tablenotemark{(b)} & 2 & 8.0 & -19.3 & 10.46 & -0.50 & Ha,24 & 0.29 & 1.6 $\pm$ 0.3 & -0.49 \\
NGC4236 & P & 8 & 4.5 & -19.2 & 9.27 & -1.00 & UV,24 & 0.01 & 1.9 $\pm$ 0.8 & -2.40 \\
NGC0247 & nb/d & 7 & 3.7 & -19.2 & 8.59 & -0.71 & UV,24 & 0.00 & ...   & ... \\
NGC6684 & C & -2 & 10.9 & -19.2 & 10.13 & -1.55 & 24 & 0.38 & 3.5 $\pm$ 0.8 & -0.49 \\
NGC1512 & P & 1 & 9.6 & -19.2 & 9.93 & -0.84 & UV,24 & 0.28 & 1.8 $\pm$ 1.2 & -1.38 \\
NGC7713 & P & 7 & 9.3 & -19.1 & 9.63 & -0.50 & Ha,24 & 0.01 & 1.1 $\pm$ 1.9 & -1.65 \\
NGC1313 & nb/d & 7 & 4.2 & -19.1 & 9.49 & -0.46 & UV,24 & 0.00 & ...   & ... \\
NGC0672 & nb/d & 6 & 7.2 & -18.9 & 9.05 & -1.11 & UV,24 & 0.00 & ...   & ... \\
NGC3377 & E & -5 & 9.3 & -18.9 & 10.47 & -2.15 & UV,24 & 1.00 & ...   & ... \\
NGC0598 & P & 6 & 0.8 & -18.9 & 9.21 & -0.91 & UV,24 & 0.03 & 1.4 $\pm$ 2.3 & -1.84 \\
NGC5068 & nb/d & 6 & 6.2 & -18.9 & 9.11 & -0.89 & UV,24 & 0.00 & ...   & ... \\
NGC3486 & P & 5 & 8.2 & -18.8 & 9.42 & -0.58 & UV,24 & 0.10 & 1.6 $\pm$ 1.1 & -2.04 \\
NGC3344 & C & 4 & 6.6 & -18.8 & 9.51 & -0.71 & UV,24 & 0.08 & 2.3 $\pm$ 0.6 & -1.59 \\
NGC7793 & P & 7 & 3.9 & -18.8 & 9.83 & -0.60 & UV,24 & 0.02 & 1.1 $\pm$ 0.8 & -1.77 \\
NGC3412 & C & -2 & 10.4 & -18.8 & 9.92 & -2.15 & UV,24 & 0.39 & 2.6 $\pm$ 0.6 & -2.39 \\
NGC6503 & P & 6 & 5.3 & -18.7 & 9.46 & -1.07 & UV,24 & 0.01 & 1.0 $\pm$ 1.5 & -1.65 \\
IC5332 & P & 7 & 9.5 & -18.7 & 9.91 & -0.37 & Ha,24 & 0.04 & 1.3 $\pm$ 0.6 & -2.12 \\
NGC1744 & nb/d & 6 & 7.7 & -18.7 & 9.43 & -0.92 & UV,24 & 0.00 & ...   & ... \\
NGC4314 & P\tablenotemark{(d)} & 1 & 9.7 & -18.6 & 9.85 & -1.16 & UV,24 & 0.22 & 3.1 $\pm$ 0.9 & -1.22 \\
NGC4605 & nb/d & 5 & 5.5 & -18.6 & 9.24 & -0.87 & UV,24 & 0.00 & ...   & ... \\
NGC4618 & P & 8 & 7.8 & -18.6 & 9.68 & -0.86 & UV,24 & 0.04 & 1.4 $\pm$ 1.8 & -1.90 \\
NGC1058 & P & 5 & 9.2 & -18.6 & 9.21 & -0.82 & UV,24 & 0.03 & 1.5 $\pm$ 0.7 & -1.79 \\
NGC1156 & nb/d & 10 & 7.8 & -18.6 & 8.92 & -0.37 & UV & 0.00 & ...   & ... \\
NGC1637 & P & 5 & 8.9 & -18.5 & 9.53 & -0.48 & 24 & 0.06 & 1.1 $\pm$ 0.4 & -0.68 \\
NGC2787 & C & 0 & 7.5 & -18.3 & 10.43 & -1.74 & UV,24 & 0.58 & 2.6 $\pm$ 0.5 & -1.87 \\
NGC3239 & nb/d & 9 & 8.3 & -18.3 & 9.25 & -0.51 & Ha & 0.00 & ...   & ... \\
UGCA90 & P & 7 & 10.4 & -18.3 & 9.43 & -0.53 & UV & 0.04 & 0.9 $\pm$ 0.9 & ... \\
NGC0024 & nb/d & 5 & 8.1 & -18.2 & 9.73 & -1.10 & UV,24 & 0.00 & ...   & ... \\
NGC4448 & P & 2 & 9.7 & -18.2 & 9.84 & -1.28 & UV,24 & 0.17 & 1.2 $\pm$ 0.9 & -1.82 \\
NGC0045 & nb/d & 8 & 7.1 & -18.2 & 9.94 & -0.44 & UV,24 & 0.00 & ...   & ... \\
NGC5474 & nb/d & 6 & 7.2 & -18.2 & 9.31 & -0.52 & UV & 0.00 & ...   & ... \\
NGC6689 & P & 6 & 9.8 & -18.2 & 9.64 & -1.03 & Ha,24 & 0.04 & 1.2 $\pm$ 1.0 & -1.32 \\
NGC0949 & P & 4 & 9.2 & -18.1 & 9.18 & 0.00 & ... & 0.20 & 1.6 $\pm$ 1.2 & ... \\
NGC6673 & P & -1 & 10.9 & -18.1 & 10.05 & ... & ... & 0.29 & 1.1 $\pm$ 0.8 & ... \\
NGC0300 & nb/d & 7 & 2.0 & -18.1 & 9.20 & -0.98 & UV,24 & 0.00 & ...   & ... \\
NGC4242 & nb/d & 8 & 7.4 & -18.1 & 10.14 & -1.24 & UV,24 & 0.00 & ...   & ... \\
LMC & nb/d & 9 & 0.1 & -18.1 & 9.20 & -0.61 & Ha & 0.00 & ...   & ... \\
NGC4136 & P & 5 & 9.7 & -18.0 & 9.01 & -1.04 & UV,24 & 0.02 & 1.8 $\pm$ 1.5 & -2.19 \\
UGCA114 & nb/d & 7 & 9.8 & -17.9 & 9.60 & -0.41 & UV & 0.00 & ...   & ... \\
NGC5585 & P & 7 & 5.7 & -17.8 & 9.28 & -1.09 & UV,24 & 0.05 & 0.9 $\pm$ 0.3 & -2.06 \\
NGC1796 & nb/d & 5 & 10.3 & -17.8 & 9.43 & -0.49 & Ha,24 & 0.00 & ...   & ... \\
NGC4245 & P & 0 & 9.7 & -17.8 & 10.21 & -1.76 & UV,24 & 0.21 & 1.0 $\pm$ 0.6 & -1.96 \\
NGC2976 & nb/d & 5 & 3.6 & -17.7 & 9.53 & -1.40 & UV,24 & 0.00 & ...   & ... \\
NGC3077 & nb/d & 6 & 3.8 & -17.7 & 9.65 & -1.12 & Ha & 0.00 & ...   & ... \\
NGC4534 & nb/d & 8 & 10.8 & -17.7 & 9.02 & -0.66 & UV & 0.00 & ...   & ... \\
NGC5102 & C & -3 & 3.4 & -17.6 & 9.25 & ... & ... & 0.37 & 3.5 $\pm$ 0.9 & ... \\
NGC5253 & nb/d & 10 & 3.2 & -17.6 & 8.87 & -0.76 & Ha & 0.00 & ...   & ... \\
NGC0959 & nb/d & 8 & 9.2 & -17.6 & 9.38 & -1.34 & UV,24 & 0.00 & ...   & ... \\
NGC1487 & M & 7 & 9.1 & -17.5 & 9.39 & -0.81 & UV,24 & 1.00 & ...   & ... \\
NGC5949 & nb/d & 4 & 8.5 & -17.4 & 9.61 & -1.51 & UV,24 & 0.00 & ...   & ... \\
NGC2552 & nb/d & 9 & 7.7 & -17.4 & 9.25 & -1.52 & UV,24 & 0.00 & ...   & ... \\
IC4710 & nb/d & 9 & 7.7 & -17.4 & 9.27 & -1.24 & UV,24 & 0.00 & ...   & ... \\
NGC4214 & P & 9 & 2.9 & -17.4 & 8.97 & -0.96 & UV,24 & 0.01 & 1.4 $\pm$ 1.1 & -1.20 \\
NGC2500 & P & 7 & 7.6 & -17.4 & 9.42 & -1.22 & UV,24 & 0.02 & 1.7 $\pm$ 1.5 & -2.13 \\
NGC4941 & P & 2 & 6.4 & -17.4 & 9.33 & -1.25 & UV,24 & 0.16 & 1.9 $\pm$ 0.7 & -1.65 \\
NGC3593 & P & 0 & 6.5 & -17.3 & 9.70 & -0.86 & UV,24 & 0.51 & 0.8 $\pm$ 0.2 & -0.94 \\
NGC4485 & nb/d & 10 & 7.1 & -17.3 & 8.76 & -0.66 & UV & 0.00 & ...   & ... \\
NGC3274 & nb/d & 8 & 9.5 & -17.3 & 8.94 & -1.13 & UV,24 & 0.00 & ...   & ... \\
NGC4020 & nb/d & 7 & 9.7 & -17.3 & 9.32 & -1.45 & UV,24 & 0.00 & ...   & ... \\
ESO305-G009 & nb/d & 8 & 10.9 & -17.2 & 8.39 & ... & ... & 0.00 & ...   & ... \\
NGC3738 & nb/d & 10 & 4.9 & -17.2 & 8.62 & -1.46 & UV,24 & 0.00 & ...   & ... \\
NGC2537 & M & 8 & 6.9 & -17.2 & 9.27 & -1.37 & UV,24 & 1.00 & ...   & -1.66 \\
UGCA212 & nb/d & 8 & 10.1 & -17.2 & 9.31 & ... & ... & 0.00 & ...   & ... \\
NGC3125 & nb/d & 10 & 10.8 & -17.1 & 8.95 & -0.43 & Ha & 0.00 & ...   & ... \\
NGC5204 & nb/d & 9 & 4.7 & -17.0 & 8.93 & -1.20 & UV,24 & 0.00 & ...   & ... \\
UGCA103 & P & 9 & 10.4 & -17.0 & 9.06 & -2.65 & Ha & 0.05 & 0.6 $\pm$ 0.5 & ... \\
NGC5408 & nb/d & 10 & 4.8 & -17.0 & 9.10 & -0.97 & Ha & 0.00 & ...   & ... \\
UGCA106 & nb/d & 9 & 9.8 & -17.0 & 9.09 & -1.17 & UV,24 & 0.00 & ...   & ... \\
NGC4625 & nb/d & 9 & 8.7 & -17.0 & 9.05 & -1.21 & UV,24 & 0.00 & ...   & ... \\
NGC2337 & nb/d & 10 & 7.9 & -17.0 & 8.58 & -1.04 & Ha & 0.00 & ...   & ... \\
NGC0855 & P & -1 & 9.7 & -17.0 & 9.50 & -1.59 & UV,24 & 0.33 & 1.2 $\pm$ 0.2 & -1.68 \\
ESO383-G087 & nb/d & 8 & 3.5 & -16.9 & 8.86 & -1.47 & Ha & 0.00 & ...   & ... \\
UGC04305 & nb/d & 10 & 3.4 & -16.9 & 8.78 & -0.84 & UV & 0.00 & ...   & ... \\
SMC & nb/d & 9 & 0.1 & -16.8 & 8.81 & -1.43 & Ha & 0.00 & ...   & ... \\
NGC1800 & nb/d & 9 & 8.2 & -16.7 & 8.71 & -1.04 & UV & 0.00 & ...   & ... \\
UGC07490 & nb/d & 9 & 8.4 & -16.7 & 8.97 & -1.61 & Ha,24 & 0.00 & ...   & ... \\
ESO435-G016 & nb/d & 3 & 9.1 & -16.7 & 9.37 & -1.12 & UV & 0.00 & ...   & ... \\
ESO158-G003 & nb/d & 9 & 10.0 & -16.7 & 8.65 & ... & Ha & 0.00 & ...   & ... \\
NGC0404 & C & -1 & 3.3 & -16.6 & 9.53 & -1.99 & UV,24 & 0.16 & 3.4 $\pm$ 1.0 & -2.07 \\
NGC3299 & nb/d & 8 & 10.4 & -16.6 & 9.54 & -1.83 & UV,24 & 0.00 & ...   & ... \\
UGC05151 & nb/d & 10 & 10.7 & -16.6 & 8.45 & -1.30 & Ha & 0.00 & ...   & ... \\
UGC07690 & nb/d & 10 & 7.7 & -16.5 & 8.74 & -1.50 & UV,24 & 0.00 & ...   & ... \\
UGC07690 & nb/d & 10 & 7.7 & -16.5 & 8.74 & -1.19 & UV & 0.00 & ...   & ... \\
NGC1510 & P & -1 & 9.8 & -16.5 & 8.86 & -1.20 & UV,24 & 0.42 & ...   & ... \\
NGC5608 & nb/d & 10 & 10.2 & -16.5 & 7.93 & -1.08 & UV & 0.00 & ...   & ... \\
ESO364-G?029 & nb/d & 10 & 7.4 & -16.5 & 8.09 & -1.60 & Ha & 0.00 & ...   & ... \\
IC5152 & nb/d & 10 & 2.0 & -16.5 & 8.59 & -1.46 & UV & 0.00 & ...   & ... \\
ESO306-G013 & nb/d & 3 & 10.8 & -16.4 & 9.24 & ... & ... & 0.00 & ...   & ... \\
IC4870 & nb/d & 10 & 9.9 & -16.4 & 8.18 & -0.89 & UV & 0.00 & ...   & ... \\
NGC4288 & nb/d & 7 & 7.7 & -16.4 & 8.75 & -1.42 & UV,24 & 0.00 & ...   & ... \\
IC5256 & nb/d & 8 & 10.8 & -16.4 & 8.06 & ... & ... & 0.00 & ...   & ... \\
NGC7518 & P & 1 & 10.0 & -16.4 & 9.25 & ... & ... & 0.16 & ...   & ... \\
UGC05451 & nb/d & 10 & 8.7 & -16.3 & 7.76 & -1.52 & UV & 0.00 & ...   & ... \\
UGC02023 & nb/d & 10 & 9.2 & -16.3 & 8.96 & -1.14 & UV & 0.00 & ...   & ... \\
UGC09660 & nb/d & 4 & 10.2 & -16.3 & 7.97 & -1.36 & UV & 0.00 & ...   & ... \\
MCG-05-13-004 & nb/d & 9 & 6.6 & -16.3 & 8.46 & ... & ... & 0.00 & ...   & ... \\
NGC4248 & nb/d & 3 & 7.2 & -16.2 & 9.20 & -2.15 & UV,24 & 0.00 & ...   & ... \\
UGC07698 & nb/d & 10 & 6.1 & -16.2 & 8.56 & -1.36 & UV & 0.00 & ...   & ... \\
UGC00891 & nb/d & 9 & 10.8 & -16.2 & 8.74 & -1.66 & UV & 0.00 & ...   & ... \\
NGC4204 & nb/d & 8 & 10.4 & -16.2 & 9.07 & -1.21 & UV,24 & 0.00 & ...   & ... \\
NGC0221 & E & -5 & 0.8 & -16.2 & 10.02 & -1.58 & UV,24 & 1.00 & 2.8 $\pm$ 0.3 & ... \\
NGC5264 & nb/d & 9 & 4.5 & -16.2 & 8.74 & -1.66 & UV & 0.00 & ...   & ... \\
UGC10736 & nb/d & 8 & 9.8 & -16.1 & 7.86 & -1.51 & UV & 0.00 & ...   & ... \\
ESO483-G013 & nb/d & -1 & 10.4 & -16.1 & 8.16 & -1.28 & UV & 0.00 & ...   & ... \\
UGC01865 & nb/d & 9 & 9.2 & -16.1 & 8.61 & -1.56 & UV & 0.00 & ...   & ... \\
UGC08201 & nb/d & 10 & 4.6 & -16.0 & 8.32 & -1.56 & UV & 0.00 & ...   & ... \\
NGC4707 & nb/d & 9 & 7.4 & -16.0 & 8.87 & -1.33 & UV & 0.00 & ...   & ... \\
UGC08188 & nb/d & 9 & 4.5 & -16.0 & 8.38 & -1.21 & UV & 0.00 & ...   & ... \\
UGC07608 & nb/d & 10 & 7.8 & -16.0 & 8.21 & -1.18 & UV & 0.00 & ...   & ... \\
MCG-03-34-002 & nb/d & 4 & 10.2 & -16.0 & 8.35 & ... & ... & 0.00 & ...   & ... \\
NGC1522 & nb/d & 10 & 9.3 & -15.9 & 8.00 & -1.11 & UV & 0.00 & ...   & ... \\
UGC05829 & nb/d & 10 & 7.9 & -15.9 & 8.17 & -0.86 & UV & 0.00 & ...   & ... \\
UGC08313 & nb/d & 5 & 8.7 & -15.9 & 7.67 & -1.61 & UV & 0.00 & ...   & ... \\
UGC01176 & nb/d & 10 & 9.0 & -15.9 & 8.79 & -1.34 & UV & 0.00 & ...   & ... \\
NGC4080 & nb/d & 10 & 6.9 & -15.9 & 7.69 & -1.59 & UV & 0.00 & ...   & ... \\
UGC02259 & nb/d & 8 & 9.2 & -15.9 & 8.01 & -1.34 & Ha & 0.00 & ...   & ... \\
ESO119-G016 & nb/d & 10 & 9.8 & -15.8 & 7.80 & -1.52 & UV & 0.00 & ...   & ... \\
NGC1705 & nb/d & 10 & 5.1 & -15.8 & 7.93 & -0.98 & UV & 0.00 & ...   & ... \\
NGC1592 & nb/d & 10 & 10.6 & -15.8 & 7.83 & 3.00 & Ha & 0.00 & ...   & ... \\
ESO435-IG020 & nb/d & 10 & 9.0 & -15.8 & 8.41 & -1.01 & Ha & 0.00 & ...   & ... \\
ESO486-G021 & nb/d & 2 & 8.9 & -15.7 & 7.79 & -1.38 & UV & 0.00 & ...   & ... \\
UGC07774 & nb/d & 7 & 7.4 & -15.7 & 7.65 & -1.84 & UV & 0.00 & ...   & ... \\
UGC06161 & nb/d & 8 & 10.3 & -15.7 & 7.64 & -1.23 & UV & 0.00 & ...   & ... \\
ESO324-G024 & nb/d & 10 & 3.7 & -15.7 & 8.55 & -1.71 & UV & 0.00 & ...   & ... \\
ESO409-IG015 & nb/d & 6 & 10.4 & -15.6 & 7.62 & -1.33 & Ha & 0.00 & ...   & ... \\
UGC05889 & nb/d & 9 & 9.3 & -15.6 & 8.68 & -2.11 & Ha & 0.00 & ...   & ... \\
UGC04426 & nb/d & 10 & 10.3 & -15.6 & 8.27 & -1.68 & UV & 0.00 & ...   & ... \\
UGC07639 & nb/d & 10 & 8.0 & -15.6 & 7.74 & -1.73 & UV & 0.00 & ...   & ... \\
UGC04787 & nb/d & 8 & 6.5 & -15.6 & 7.37 & -1.86 & UV & 0.00 & ...   & ... \\
UGC05672 & nb/d & 5 & 6.3 & -15.5 & 7.44 & -2.20 & UV & 0.00 & ...   & ... \\
ESO245-G005 & nb/d & 10 & 4.4 & -15.5 & 9.00 & -1.24 & UV & 0.00 & ...   & ... \\
UGC01104 & nb/d & 9 & 7.5 & -15.5 & 8.03 & -1.59 & UV & 0.00 & ...   & ... \\
UGC07719 & nb/d & 8 & 9.4 & -15.5 & 7.52 & -1.60 & UV & 0.00 & ...   & ... \\
UGC01056 & nb/d & 10 & 10.3 & -15.5 & 8.11 & -1.73 & Ha & 0.00 & ...   & ... \\
ESO302-G014 & nb/d & 10 & 9.6 & -15.5 & 7.29 & -1.48 & UV & 0.00 & ...   & ... \\
UGC09405 & nb/d & 10 & 8.0 & -15.5 & 8.09 & -2.04 & UV & 0.00 & ...   & ... \\
UGC06457 & nb/d & 10 & 10.2 & -15.4 & 7.61 & -1.73 & UV & 0.00 & ...   & ... \\
NGC5477 & nb/d & 9 & 7.7 & -15.4 & 7.88 & -1.38 & UV & 0.00 & ...   & ... \\
ISZ399 & nb/d & 10 & 9.0 & -15.4 & 7.88 & -1.50 & Ha & 0.00 & ...   & ... \\
ESO059-G001 & nb/d & 10 & 4.6 & -15.3 & 8.67 & -2.19 & Ha & 0.00 & ...   & ... \\
UGC07267 & nb/d & 8 & 7.3 & -15.3 & 7.35 & -1.99 & UV & 0.00 & ...   & ... \\
UGC05923 & nb/d & 0 & 7.2 & -15.3 & 7.69 & -2.05 & UV & 0.00 & ...   & ... \\
ESO383-G091 & nb/d & 7 & 3.6 & -15.3 & 8.01 & -3.12 & Ha & 0.00 & ...   & ... \\
NGC4068 & nb/d & 10 & 4.3 & -15.2 & 7.67 & -1.87 & Ha & 0.00 & ...   & ... \\
ESO381-G020 & nb/d & 10 & 5.4 & -15.2 & 7.76 & -1.70 & UV & 0.00 & ...   & ... \\
UGC04115 & nb/d & 10 & 7.7 & -15.2 & 7.40 & -1.63 & UV & 0.00 & ...   & ... \\
UGC01561 & nb/d & 10 & 10.5 & -15.2 & 7.81 & -1.44 & UV & 0.00 & ...   & ... \\
ESO377-G003 & nb/d & 4 & 9.2 & -15.2 & 7.83 & ... & ... & 0.00 & ...   & ... \\
UGC09497 & nb/d & 6 & 10.0 & -15.1 & 7.42 & -1.67 & UV & 0.00 & ...   & ... \\
UGC12713 & nb/d & 0 & 7.7 & -15.1 & 8.09 & -1.92 & UV & 0.00 & ...   & ... \\
UGC07949 & nb/d & 10 & 9.9 & -15.1 & 7.41 & -1.59 & UV & 0.00 & ...   & ... \\
ESO149-G003 & nb/d & 10 & 6.4 & -15.1 & 7.36 & -1.79 & UV & 0.00 & ...   & ... \\
IC4247 & nb/d & 2 & 5.0 & -15.1 & 7.80 & -2.13 & UV & 0.00 & ...   & ... \\
CGCG262-028 & nb/d & 5 & 6.9 & -15.1 & 7.95 & -1.64 & UV & 0.00 & ...   & ... \\
UGC05692 & nb/d & 9 & 4.0 & -15.1 & 8.52 & -2.31 & UV & 0.00 & ...   & ... \\
UGC03860 & nb/d & 10 & 7.8 & -15.0 & 7.93 & -1.79 & UV & 0.00 & ...   & ... \\
UGC06900 & nb/d & 10 & 7.5 & -15.0 & 8.64 & -2.11 & UV & 0.00 & ...   & ... \\
UGCA319 & nb/d & 9 & 7.4 & -15.0 & 8.03 & -2.03 & UV & 0.00 & ...   & ... \\
IC2782 & nb/d & 8 & 9.7 & -15.0 & 7.50 & ... & ... & 0.00 & ...   & ... \\
UGC07271 & nb/d & 7 & 7.8 & -15.0 & 7.21 & -1.95 & UV & 0.00 & ...   & ... \\
UGC05456 & nb/d & 5 & 3.8 & -15.0 & 8.00 & -1.92 & UV & 0.00 & ...   & ... \\
UGC03966 & nb/d & 10 & 6.8 & -15.0 & 8.18 & -1.76 & UV & 0.00 & ...   & ... \\
UGC07866 & nb/d & 10 & 4.6 & -14.9 & 7.73 & -1.66 & UV & 0.00 & ...   & ... \\
SBS1331+493 & nb/d & 10 & 9.3 & -14.9 & 6.76 & -1.85 & Ha & 0.00 & ...   & ... \\
UGC00695 & nb/d & 6 & 10.2 & -14.9 & 7.46 & -1.74 & UV & 0.00 & ...   & ... \\
UGC02014 & nb/d & 10 & 9.2 & -14.9 & 7.72 & -2.38 & UV & 0.00 & ...   & ... \\
ESO104-G044 & nb/d & 9 & 8.4 & -14.9 & 7.57 & -1.75 & Ha & 0.00 & ...   & ... \\
NGC5238 & nb/d & 8 & 5.2 & -14.9 & 7.51 & -1.77 & UV & 0.00 & ...   & ... \\
UGC04998 & nb/d & 10 & 10.5 & -14.9 & 8.16 & -1.89 & UV & 0.00 & ...   & ... \\
NGC5229 & nb/d & 7 & 5.1 & -14.9 & 8.18 & -1.94 & UV & 0.00 & ...   & ... \\
UGC07916 & nb/d & 10 & 8.2 & -14.9 & 7.34 & -1.60 & UV & 0.00 & ...   & ... \\
UGCA153 & nb/d & 10 & 6.5 & -14.9 & 7.84 & -2.05 & UV & 0.00 & ...   & ... \\
UGCA298 & nb/d & -1 & 10.3 & -14.9 & 7.15 & ... & ... & 0.00 & ...   & ... \\
UGC07599 & nb/d & 8 & 6.9 & -14.9 & 7.85 & -1.89 & UV & 0.00 & ...   & ... \\
UGC09893 & nb/d & 7 & 10.9 & -14.9 & 7.28 & -1.71 & UV & 0.00 & ...   & ... \\
ESO249-G036 & nb/d & 10 & 9.6 & -14.8 & 7.05 & -1.64 & UV & 0.00 & ...   & ... \\
UGC05453 & nb/d & 10 & 9.3 & -14.8 & 7.29 & -2.80 & Ha & 0.00 & ...   & ... \\
UGC05288 & nb/d & 8 & 6.8 & -14.8 & 7.65 & -1.76 & UV & 0.00 & ...   & ... \\
UGC05139 & nb/d & 10 & 3.8 & -14.8 & 7.56 & -1.81 & UV & 0.00 & ...   & ... \\
ESO104-G022 & nb/d & 10 & 8.7 & -14.8 & 7.96 & -1.82 & Ha & 0.00 & ...   & ... \\
KUG1004+392 & nb/d & 10 & 7.8 & -14.8 & 7.10 & -1.88 & UV & 0.00 & ...   & ... \\
UGC07577 & nb/d & 10 & 2.7 & -14.8 & 8.00 & -2.19 & UV & 0.00 & ...   & ... \\
UGC02716 & nb/d & 8 & 6.2 & -14.7 & 8.67 & -1.97 & UV & 0.00 & ...   & ... \\
UGC05740 & nb/d & 9 & 9.3 & -14.7 & 7.29 & -1.55 & UV & 0.00 & ...   & ... \\
UGC07678 & nb/d & 6 & 9.3 & -14.7 & 6.88 & -1.37 & UV & 0.00 & ...   & ... \\
UGC03817 & nb/d & 10 & 8.6 & -14.7 & 8.05 & -2.05 & Ha & 0.00 & ...   & ... \\
UGC07559 & nb/d & 10 & 4.9 & -14.7 & 7.65 & -1.79 & UV & 0.00 & ...   & ... \\
UGC09992 & nb/d & 10 & 8.6 & -14.7 & 7.70 & -1.84 & UV & 0.00 & ...   & ... \\
UGC07950 & nb/d & 10 & 7.9 & -14.6 & 7.20 & -1.48 & UV & 0.00 & ...   & ... \\
CGCG217-018 & nb/d & 10 & 8.2 & -14.6 & 7.14 & -1.89 & UV & 0.00 & ...   & ... \\
UGC07242 & nb/d & 6 & 5.4 & -14.6 & 7.26 & ... & ... & 0.00 & ...   & ... \\
ESO325-G011 & nb/d & 10 & 3.4 & -14.6 & 7.88 & -1.91 & Ha & 0.00 & ...   & ... \\
UGC09211 & nb/d & 10 & 10.7 & -14.6 & 7.75 & -1.39 & UV & 0.00 & ...   & ... \\
MCG-04-02-003 & nb/d & 9 & 9.8 & -14.6 & 7.20 & -2.20 & Ha & 0.00 & ...   & ... \\
ESO252-IG001 & nb/d & 99 & 6.0 & -14.5 & 7.62 & -2.42 & Ha & 0.00 & ...   & ... \\
UGC05797 & nb/d & 10 & 6.8 & -14.5 & 7.24 & -2.14 & UV & 0.00 & ...   & ... \\
UGCA281 & nb/d & 10 & 5.7 & -14.5 & 6.82 & -1.41 & Ha & 0.00 & ...   & ... \\
UGC05423 & nb/d & 10 & 5.3 & -14.4 & 7.75 & -2.33 & UV & 0.00 & ...   & ... \\
UGC05373 & nb/d & 10 & 1.4 & -14.4 & 7.90 & -2.29 & UV & 0.00 & ...   & ... \\
UGC05427 & nb/d & 8 & 7.1 & -14.3 & 6.90 & -1.92 & UV & 0.00 & ...   & ... \\
UGC06102 & nb/d & 10 & 8.5 & -14.3 & 7.22 & -1.89 & UV & 0.00 & ...   & ... \\
ESO553-G046 & nb/d & 1 & 5.0 & -14.3 & 7.39 & ... & ... & 0.00 & ...   & ... \\
KUG1413+573 & nb/d & 10 & 7.4 & -14.3 & 7.03 & -2.32 & UV & 0.00 & ...   & ... \\
UGC00685 & nb/d & 9 & 4.7 & -14.3 & 7.62 & -2.19 & UV & 0.00 & ...   & ... \\
UGC05076 & nb/d & 10 & 8.3 & -14.3 & 7.15 & -2.91 & Ha & 0.00 & ...   & ... \\
ESO140-G019 & nb/d & 10 & 10.8 & -14.3 & 7.63 & -1.75 & Ha & 0.00 & ...   & ... \\
UGC05918 & nb/d & 10 & 7.4 & -14.3 & 7.67 & -1.95 & UV & 0.00 & ...   & ... \\
UGC08024 & nb/d & 10 & 4.3 & -14.3 & 7.52 & -1.81 & UV & 0.00 & ...   & ... \\
UGC08683 & nb/d & 10 & 9.6 & -14.3 & 7.00 & -1.57 & UV & 0.00 & ...   & ... \\
UGC00668 & nb/d & 10 & 0.7 & -14.2 & 8.00 & -2.09 & UV & 0.00 & ...   & ... \\
ESO238-G005 & nb/d & 10 & 8.9 & -14.2 & 7.01 & -1.81 & UV & 0.00 & ...   & ... \\
NGC6789 & nb/d & 10 & 3.6 & -14.2 & 7.36 & -2.43 & Ha & 0.00 & ...   & ... \\
IC4316 & nb/d & 10 & 4.4 & -14.2 & 7.89 & ... & ... & 0.00 & ...   & ... \\
MRK36 & nb/d & 10 & 7.8 & -14.1 & 7.09 & -1.43 & Ha & 0.00 & ...   & ... \\
UGC09240 & nb/d & 10 & 2.8 & -14.1 & 7.46 & -2.24 & UV & 0.00 & ...   & ... \\
ESO300-G016 & nb/d & 10 & 7.8 & -14.1 & 7.01 & -2.28 & UV & 0.00 & ...   & ... \\
UGC06541 & nb/d & 10 & 3.9 & -14.1 & 7.36 & -2.08 & Ha & 0.00 & ...   & ... \\
NGC4190 & nb/d & 10 & 3.5 & -14.0 & 7.08 & -1.99 & UV & 0.00 & ...   & ... \\
AM0704-582 & nb/d & 9 & 4.9 & -14.0 & 7.62 & -2.23 & Ha & 0.00 & ...   & ... \\
UGC02684 & nb/d & 10 & 6.5 & -13.9 & 8.12 & -2.19 & UV & 0.00 & ...   & ... \\
ESO473-G024 & nb/d & 10 & 8.0 & -13.9 & 6.98 & -2.05 & UV & 0.00 & ...   & ... \\
IC2787 & nb/d & 6 & 7.7 & -13.9 & 6.87 & -4.84 & Ha & 0.00 & ...   & ... \\
MCG+07-26-012 & nb/d & 6 & 6.4 & -13.9 & 6.72 & -3.07 & Ha & 0.00 & ...   & ... \\
ESO348-G009 & nb/d & 10 & 8.6 & -13.9 & 6.88 & -1.77 & UV & 0.00 & ...   & ... \\
AM0605-341 & nb/d & 10 & 7.0 & -13.9 & 7.38 & -1.98 & Ha & 0.00 & ...   & ... \\
UGC07007 & nb/d & 9 & 10.1 & -13.8 & 6.93 & -1.90 & UV & 0.00 & ...   & ... \\
UGC05336 & nb/d & 10 & 3.7 & -13.8 & 7.51 & -1.99 & UV & 0.00 & ...   & ... \\
UGCA290 & nb/d & 10 & 6.7 & -13.8 & 6.86 & -1.83 & UV & 0.00 & ...   & ... \\
IC0559 & nb/d & 5 & 4.9 & -13.8 & 7.03 & -2.37 & UV & 0.00 & ...   & ... \\
ESO444-G084 & nb/d & 10 & 4.6 & -13.7 & 7.20 & -2.21 & UV & 0.00 & ...   & ... \\
MCG+07-26-011 & nb/d & 8 & 6.0 & -13.7 & 6.64 & -2.76 & Ha & 0.00 & ...   & ... \\
UGC08245 & nb/d & 10 & 3.6 & -13.7 & 6.89 & -2.60 & UV & 0.00 & ...   & ... \\
UGC07605 & nb/d & 10 & 4.4 & -13.7 & 6.92 & -2.16 & UV & 0.00 & ...   & ... \\
ESO384-G016 & nb/d & 10 & 4.5 & -13.7 & 7.36 & -4.66 & Ha & 0.00 & ...   & ... \\
UGC05917 & nb/d & 10 & 10.3 & -13.7 & 7.17 & -1.74 & UV & 0.00 & ...   & ... \\
MRK475 & nb/d & 10 & 9.0 & -13.7 & 7.16 & -1.62 & Ha & 0.00 & ...   & ... \\
UGC10669 & nb/d & 10 & 9.2 & -13.7 & 6.99 & -2.53 & UV & 0.00 & ...   & ... \\
UGC08638 & nb/d & 10 & 4.3 & -13.6 & 6.67 & -2.25 & UV & 0.00 & ...   & ... \\
UGC06456 & nb/d & 10 & 4.3 & -13.6 & 6.75 & -1.92 & UV & 0.00 & ...   & ... \\
SextansA & nb/d & 10 & 1.3 & -13.6 & 7.52 & -1.92 & UV & 0.00 & ...   & ... \\
UGC07584 & nb/d & 9 & 7.3 & -13.6 & 6.84 & -2.22 & UV & 0.00 & ...   & ... \\
CGCG035-007 & nb/d & 5 & 5.2 & -13.5 & 7.51 & -2.51 & UV & 0.00 & ...   & ... \\
UGC12894 & nb/d & 10 & 8.2 & -13.5 & 7.63 & ... & ... & 0.00 & ...   & ... \\
UGC04459 & nb/d & 10 & 3.6 & -13.4 & 7.21 & -2.16 & Ha & 0.00 & ...   & ... \\
NGC3741 & nb/d & 10 & 3.2 & -13.4 & 6.89 & -2.23 & Ha & 0.00 & ...   & ... \\
UGC08651 & nb/d & 10 & 3.0 & -13.4 & 6.63 & -2.42 & UV & 0.00 & ...   & ... \\
NGC4163 & nb/d & 10 & 3.0 & -13.3 & 6.99 & -2.34 & UV & 0.00 & ...   & ... \\
UGC08308 & nb/d & 10 & 4.2 & -13.2 & 6.52 & -2.42 & UV & 0.00 & ...   & ... \\
KUG1207+367 & nb/d & 10 & 4.5 & -13.1 & 6.76 & -2.71 & UV & 0.00 & ...   & ... \\
UGC07356 & nb/d & 10 & 6.7 & -13.1 & 6.88 & ... & ... & 0.00 & ...   & ... \\
LSBCD564-08 & nb/d & 10 & 8.7 & -13.1 & 6.71 & -4.43 & Ha & 0.00 & ...   & ... \\
UGC08508 & nb/d & 10 & 2.7 & -13.1 & 7.11 & -2.77 & Ha & 0.00 & ...   & ... \\
UGC04483 & nb/d & 10 & 3.2 & -13.0 & 6.68 & -2.41 & UV & 0.00 & ...   & ... \\
LSBCD565-06 & nb/d & 10 & 9.1 & -13.0 & 6.88 & -4.79 & Ha & 0.00 & ...   & ... \\
KDG61 & nb/d & 8 & 3.6 & -12.9 & 8.08 & -3.21 & UV & 0.00 & ...   & ... \\
AndIV & nb/d & 10 & 6.1 & -12.9 & 6.93 & -2.52 & UV & 0.00 & ...   & ... \\
LEDA166137 & nb/d & 10 & 6.0 & -12.9 & 6.42 & -2.41 & UV & 0.00 & ...   & ... \\
UGC08055 & nb/d & 10 & 6.6 & -12.9 & 6.45 & -2.17 & UV & 0.00 & ...   & ... \\
BTS76 & nb/d & 10 & 6.0 & -12.9 & 6.58 & -3.63 & UV & 0.00 & ...   & ... \\
UGC05209 & nb/d & 10 & 6.4 & -12.8 & 6.59 & -2.76 & UV & 0.00 & ...   & ... \\
UGCA438 & nb/d & 10 & 2.2 & -12.8 & 6.25 & -2.57 & UV & 0.00 & ...   & ... \\
LEDA100404 & nb/d & 9 & 6.8 & -12.7 & 6.73 & -4.92 & Ha & 0.00 & ...   & ... \\
UGC07298 & nb/d & 10 & 4.2 & -12.7 & 6.45 & ... & ... & 0.00 & ...   & ... \\
UGC09128 & nb/d & 10 & 2.2 & -12.7 & 6.66 & -2.98 & UV & 0.00 & ...   & ... \\
KKH34 & nb/d & 10 & 4.6 & -12.6 & 8.54 & -3.95 & Ha & 0.00 & ...   & ... \\
UGC08091 & nb/d & 10 & 2.1 & -12.6 & 6.68 & -2.58 & UV & 0.00 & ...   & ... \\
CGCG269-049 & nb/d & 10 & 3.2 & -12.4 & 6.54 & -2.80 & UV & 0.00 & ...   & ... \\
UGC08833 & nb/d & 10 & 3.2 & -12.4 & 6.35 & -2.75 & UV & 0.00 & ...   & ... \\
UGC05428 & nb/d & 10 & 3.5 & -12.4 & 7.42 & -4.58 & Ha & 0.00 & ...   & ... \\
UGC08215 & nb/d & 10 & 4.6 & -12.4 & 6.28 & -2.88 & UV & 0.00 & ...   & ... \\
LSBCF573-01 & nb/d & 10 & 7.2 & -12.4 & 6.35 & -4.69 & Ha & 0.00 & ...   & ... \\
LSBCD634-03 & nb/d & 10 & 9.5 & -12.1 & 6.60 & -4.45 & Ha & 0.00 & ...   & ... \\
SDSSJ0825+3532 & nb/d & 10 & 9.3 & -12.0 & 6.66 & -2.25 & Ha & 0.00 & ...   & ... \\
ESO349-G031 & nb/d & 10 & 3.2 & -12.0 & 6.14 & -5.10 & Ha & 0.00 & ...   & ... \\
KKH37 & nb/d & 10 & 3.4 & -12.0 & 6.70 & -3.10 & UV & 0.00 & ...   & ... \\
UGCA276 & nb/d & 10 & 3.2 & -11.9 & 6.17 & ... & ... & 0.00 & ...   & ... \\
UGC12613 & nb/d & 10 & 0.8 & -11.9 & 7.60 & -3.49 & UV & 0.00 & ...   & ... \\
UKS1424-460 & nb/d & 10 & 3.6 & -11.8 & 6.51 & -3.59 & Ha & 0.00 & ...   & ... \\
UGCA292 & nb/d & 10 & 3.1 & -11.8 & 6.04 & -2.76 & UV & 0.00 & ...   & ... \\
DDO210 & nb/d & 10 & 0.9 & -11.6 & 6.26 & -3.80 & UV & 0.00 & ...   & ... \\
UGC05364 & nb/d & 10 & 0.7 & -11.6 & 6.48 & -3.28 & UV & 0.00 & ...   & ... \\
UGC05272b & nb/d & 10 & 7.1 & -11.6 & 6.45 & -2.80 & UV & 0.00 & ...   & ... \\
UGCA20 & nb/d & 10 & 9.0 & -11.4 & 7.56 & -1.77 & UV & 0.00 & ...   & ... \\
KDG73 & nb/d & 10 & 3.7 & -11.0 & 5.69 & -5.47 & Ha & 0.00 & ...   & ... \\
LEDA166115 & nb/d & -1 & 4.5 & -9.8 & 6.36 & -5.41 & Ha & 0.00 & ...   & ... \\
ESO245-G007 & nb/d & 10 & 0.4 & -9.7 & 6.36 & ... & ... & 0.00 & ...   & ... \\
BK3N & nb/d & 10 & 4.0 & -9.6 & 5.75 & -5.46 & Ha & 0.00 & ...   & ... \\
M81dwA & nb/d & 10 & 3.6 & -9.2 & 5.47 & -5.20 & Ha & 0.00 & ...   & ... \\
KKR03 & nb/d & 10 & 2.1 & -8.9 & 6.59 & -5.68 & Ha & 0.00 & ...   & ... \\
LGS3 & nb/d & 99 & 0.6 & -7.9 & 5.94 & -6.90 & Ha & 0.00 & ...   & ... \\
LeoT & nb/d & 10 & 0.4 & -6.9 & 4.57 & -5.92 & Ha & 0.00 & ...   & ... \\
\tablenotetext{(a)}{E -- Elliptical Galaxy; C -- Classical bulge; P --
  Pseudobulge; nb/d -- no bulge/dwarf; M -- advanced stage merger}
\tablenotetext{(b)}{NGC~5194 \& NGC~5195 are currently interacting.}
\tablenotetext{(c)}{Categorized as classical bulge due to S\'ersic
  index despite nuclear morphology}
\tablenotetext{(d)}{Morphology strongly indicates pseudobulge, despite
  high S\'ersic index}  
\enddata
\end{deluxetable}
\clearpage
\end{landscape}
\end{center}

\end{document}